\documentclass[11pt,english]{elsart}

\usepackage{amssymb}
\usepackage{graphicx}

\usepackage[latin1]{inputenc}
\usepackage[british]{babel}
\usepackage[T1]{fontenc}

\begin{document}

    \begin{frontmatter}
    \date{}
    \title{The Sunyaev-Zeldovich MITO Project}
    \author[Rome]{L. D'Alba}%\ead{livia.dalba@roma1.infn.it},
    \author[Rome]{F. Melchiorri},
    \author[Rome]{M. De Petris},
    \author[Rome]{A. Orlando},
    \author[Rome]{L. Lamagna},
    \author[TelAviv]{Y. Rephaeli},
    \author[Monteporzio]{S. Colafrancesco},
    \author[Paris]{M. Signore},
    \author[Bonn]{E. Kreysa}
    \address[Rome]{Dipartimento di Fisica, Università "La Sapienza", Roma,
    Italy}
    \address[TelAviv]{Tel Aviv University, Tel Aviv, Israel}
    \address[Monteporzio]{Osservatori di Roma, Monteporzio (RM), Italy}
    \address[Paris]{Observatoire de Paris, Paris, France}
    \address[Bonn]{Max Planck Institut für Radioastronomie, Bonn, Germany}

        \begin{abstract}
        Compton scattering of the cosmic microwave background radiation
        by electrons in the hot gas in clusters of galaxies - the
        Sunyaev-Zeldovich (S-Z) effect -  has long been recognized as a
        uniquely important feature, rich in cosmological and
        astrophysical information. We briefly describe the effect,
        and emphasize the need for detailed S-Z and X-ray measurements
        of nearby clusters in order to use the effect as a precise
        cosmological probe. This is the goal of the MITO project,
        whose first stage consisted of observations of the S-Z effect in
        the Coma cluster. We report the results of these observations.
        \end{abstract}


    \end{frontmatter}

%\newpage

\section{Introduction}
\label{sec:introduction}

The Sunyaev-Zeldovich (S-Z) effect arises from Compton scattering
of the cosmic microwave background (CMB) radiation by electrons
in the hot gas in clusters of galaxies (for reviews, see
Rephaeli, 1995 and Birkinshaw, 1999). The fact that the effect is
essentially independent of redshift makes it an extremely
valuable cosmological probe. Interest in this effect heightened
particularly when high quality images of the effect were obtained
with interferometric arrays equipped with sensitive receivers.
More than twenty clusters have already been mapped by the BIMA
and OVRO arrays (Carlstrom et al., 1999). Together with the much
improved X-ray spectral and spatial measurements of clusters,
achieved with the \emph{Chandra} and \emph{XXM} satellites, it
seems quite likely that we will soon be much closer to realizing
the full potential of the effect as an indispensable cosmological
probe.

Multi-frequency S-Z measurements, combined with high quality
spectral and spatial X-ray measurements, will likely yield exact
values determination of the cluster baryon fraction, $\Omega_M$,
the peculiar velocity, the cluster angular diameter distance,
$d_A$, and of the Hubble constant, $H_0$. These quantities can be
accurately determined only if we have a precise description of
the intracluster (IC) gas temperature and density profiles, and
by a much improved control of systematic errors. In spite of the
major progress expected from measurements of clusters with the
new \emph{Chandra} and \emph{XXM} satellites, the substantial
uncertainties in modeling the gas thermal and spatial
distributions strongly favor use of nearby clusters for which our
knowledge of these distributions is optimal.

In this paper we describe the Sunyaev-Zeldovich effect, and
report preliminary results of observations of the Coma cluster
with the MITO experiment.

\section{Theory: the \emph{thermal }and the \emph{kinematic }S-Z effect}
\label{sec:theory}

The original description of the effect (Sunyaev \& Zeldovich 1972)
was based on a non-relativistic treatment, but an exact
description necessitates a relativistic treatment (Rephaeli,
1995). In the non-relativistic (first order) limit, the effect
consists of two distinct components which can be calculated
separately.

\begin{figure}[htpb]
  \centering
  \includegraphics[width=7cm,height=10cm]{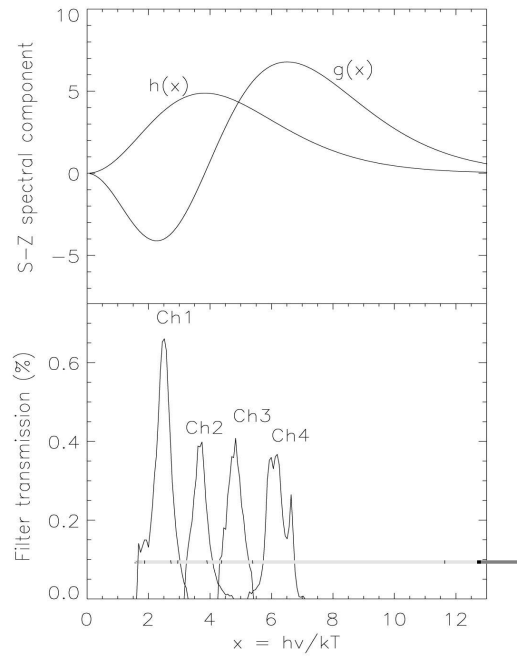}
  \caption{Spectral signature of thermal and kinematic components of the
    S-Z effect and transmission profiles of FotoMITO channels}
    \label{fig:SZcomponents}
\end{figure}

The thermal component of the S-Z effect arises from the thermal
motion of the electrons: the energy transfer from the hot gas to
the radiation produces a distortion of the CMB spectrum resulting
in a frequency shift toward the Wien side of the spectrum. Thus,
it will be a decrement in brightness at low frequencies and an
increment in brightness at high frequencies. In a
non-relativistic treatment, the CMB intensity change along the
line of sight is given by (Zeldovich and Sunyaev, 1968)
\begin{equation}\label{eq:intensity}
\Delta{I}_{nr} = I_0\,y\,g(x)
\end{equation}
where $x\,=\,\frac{h\nu}{kT_0}$, $T_0$ is the CMB temperature,
$g(x)$ expresses the spectral dependence, $y$ is the
Comptonization parameter, and
\begin{equation}\label{eq:intensity0}
I_0 = \frac{2(kT_0)^3}{(hc)^2}.
\end{equation}
The spectral form of the thermal component is expressed in the
function
\begin{equation}\label{eq:thermalspectrum}
g(x) = \frac{x^4e^x}{(e^x - 1)^2}\left[\frac{x(e^x + 1)}{e^x - 1}
- 4 \right]
\end{equation}
which vanishes at the crossover frequency
$x_0\,=\,3.83\,(\nu_0\,=\,217$ GHz) for $T_0 = 2.726K$ (see
Fig.\ref{fig:SZcomponents}). The spatial dependence is provided by
the Comptonization parameter,
\begin{equation}\label{eq:Compton}
y = \int n_{e}(r)\sigma_{T}\frac{k_{B}T_{e}(r)}{m_{e}c^2}\,dl
\end{equation}
where the integral is calculated over a line of sight through the
cluster, $n_e$ and $T_e$ are the electron density and temperature,
and $\sigma_{T}$ is the Thomson cross section. Finally, the
thermodynamic temperature change due to the scattering is
\begin{equation}\label{eq:deltaT}
\Delta{T}_{nr} = \left[\frac{x(e^x + 1)}{e^x - 1} - 4
\right]T_0\,y.
\end{equation}
Electron velocities in the hot IC gas are very high, thus
necessitating an exact relativistic treatment (Rephaeli, 1995).
Use of the relativistically correct expression for the intensity
change is essential at high frequencies, especially when the
effect is used as a cosmological probe.

The kinematic component, due to the cluster velocity of the IC
plasma with respect to the CMB rest frame, produces an additional
kinematic intensity change. Scattered photons suffer a Doppler
shift dependent on the angle of their scattering relative to the
bulk velocity. The resulting CMB brightness change is
\begin{equation}\label{}
 \Delta{I}_K = -I_0\,h(x)\frac{V_r}{c}\,\tau
\end{equation}
where $V_r$ is the line of sight component of the peculiar
velocity, and $\tau$ is the cluster optical depth given by
\begin{equation}\label{eq:depth}
\tau = \int n_{e}(r)\sigma_{T}\,dl.
\end{equation}
The spectral character of the kinematic effect is given by the
function $h(x)$
\begin{equation}\label{eq:kinematicspectrum}
h(x) = \frac{x^4e^x}{(e^x - 1)^2}
\end{equation}
Finally, the expression of the temperature change due to the
kinematic component is
\begin{equation}\label{eq:kineticdeltaK}
\Delta{T}_K = -T_0\frac{V_r}{c}\tau
\end{equation}
which is frequency independent.

Observational separation between the thermal and kinematic
effects is realistically feasible only at millimeter wavelengths,
close to the crossover frequency, as can be seen in
Fig.\ref{fig:SZcomponents}.

\section{Combined analysis of S-Z and X-ray measurements}
\label{sec:S-Z signal}

Basic parameters which characterize the state of IC gas are the
electron number density, $n_e(r)$, and the gas temperature,
$T_e(r)$. Clearly, the scattering optical depth and the
Comptonization parameter depend on these parameters. Also the
cluster X-ray spectral surface brightness can be expressed as a
function of gas density and temperature profiles

\begin{equation}
b_{X}(E) = \frac{1}{4\pi(1+z)^3}\int n_{e}(r)^2\Lambda(E,T_{e})\
dl \label{eq:brightness}
\end{equation}

where z is the redshift of the cluster and $\Lambda$ is the
spectral emissivity of the gas at an energy E. The factor $4\pi$
arises from the assumption that the emissivity is isotropic,
while the $(1+z)^3$ factor takes account of the cosmological
transformations of spectral surface brightness and energy.

The X-ray satellites ROSAT and ASCA have observed a large number
of clusters with high signal-to-noise ratios in order to
determine the gas thermal and spatial distributions: the spectrum
of emission can be used to determine the gas temperature, while
the angular distribution of X-ray emission can be used to
determine the density profile. The prediction, however, can not
be very accurate, since there is no unique inversion of equation
(\ref{eq:brightness}), necessitating use of a parameterized model
to fit the X-ray data. The density and temperature distributions
are commonly described by the \emph{isothermal $\beta$ model}
(Cavaliere, Fusco-Femiano, 1976, 1978). In this simplified
description the isothermal gas is assumed to be spherically
distributed

\begin{equation}
n_{e}(r)=n_{e0}\left(1+\frac{r^2}{r_c^2}\right)^{-\frac{3}{2}\beta}
\end{equation}

where $r_c$ is the gas core radius. Deviations from this model
have been noted, especially in cluster outer regions. For a
recent discussion on the influence of the shape and the finite
extension of a cluster on both the X-ray surface brightness and
the S-Z effect see Puy et al., 2000 and references therein.

Combining measurements of the X-ray emissivity $(\propto{n_e^2})$
and of the thermal S-Z effect $(\propto{n_e})$, a substantial
part of the indeterminacy inherent in the X-ray analysis of IC gas
properties can be removed. This is very important, since the
accuracy in the determination of the appropriate model describing
IC gas has important ramifications on the accuracy in the
determination of the Hubble constant, $H_0$. This cosmological
parameter can be estimated comparing the theoretical expression
of the angular diameter distance, $d_A$, to a cluster, with its
measured value. In fact, assuming spherical symmetry and using
the cluster's angular size, the absolute distance to a cluster
can be calculated without recourse to the cosmic distance ladder.
To yield this result, we need only to know the emission of the IC
gas, which is provided by the X-ray surface brightness, and its
absorption, which is provided by the thermal S-Z effect
(Holzapfel et al., 1997). This observational value must be
compared with the theoretic expression for $d_A$ given by (Goobar
and Perlmutter, 1995)
\newline
\begin{flushleft}
$H_0\,d_A(z,\Omega_M,\Omega_\Lambda) = $ \\
\end{flushleft}
\begin{equation}
=\,\frac{c}{(1+z)\sqrt{|\kappa|}}S\left(\sqrt{|\kappa|}\int_0^z[(1+z')^2(1+\Omega_Mz')-z'(
2+z')\Omega_\Lambda]^{-\frac{1}{2}}\ dz'\right)
\end{equation}

where $\Omega_M$ and $\Omega_\Lambda$ are the mass density and
the cosmological constant density parameters respectively, and
$\kappa = 1-\,\Omega_M\,-\,\Omega_\Lambda$ is the curvature
parameter. For $\Omega_M\,+\,\Omega_\Lambda\,>\,1$, $S(x)$ is
defined as $\sin{(x)}$ and $\kappa\,<\,0$; for
$\Omega_M\,+\,\Omega_\Lambda\,<\,1$, $S(x)\,=\,\sinh{(x)}$ and
$\kappa\,>\,0$; and for $\Omega_M\,+\,\Omega_\Lambda\,=\,1$,
$S(x)\,=\,x$ and $\kappa\,=\,0$.

\section{Observing nearby clusters}
\label{sub:distant}

The S-Z effect, as just discussed, can be used as a distance scale
indicator. Even though the S-Z effect is independent of distance,
the quality of observational data favors observations of nearby
clusters for use of the effect as a cosmological probe. The main
advantage of observing distant clusters in the mm region is the
reduced contamination from the primary CMB anisotropy; however,
the quality of X-ray spectral and spatial data is significantly
lower (due to geometric dimming, and limited spatial resolution)
than that available for nearby clusters. Non accurate clusters
X-ray images cause a large uncertainty in determining the
parameters of the \emph{isothermal $\beta$ model}, and thus
inaccurate values of the Hubble constant. The large systematic
errors that affect X-ray data can be limited observing nearby
clusters for which better X-ray maps are available. This allows
to model more accurately the structure of IC gas. Obviously, to
resolve the S-Z effect in nearby clusters, all the problems
relative to the detection of millimetric signals must be solved ,
first of all the atmospheric noise but also the foreground
confusion caused by other astrophysical sources.

In the review of Birkinshaw (Birkinshaw, 1999) there is a
complete list (since 1999) of the experiments realized in order
to detect the S-Z effect toward nearby and distant clusters of
galaxies. For each experiment, the relative measured CMB
temperature change is reported. In the following section we
describe the new MITO experiment and project, whose main goal is
the measurement of the S-Z effect in nearby clusters, beginning
with the Coma cluster.

\section{MITO: Millimetric and Infrared Testagrigia Observatory}
\label{sec:Mito}

MITO is a ground-based observatory located at 3480 m a.s.l., on
the top of the Testa Grigia mountain, in Val d'Aosta, Italy (M. De
Petris et al.,1996). Its observational range is the mm and submm
region of the spectrum, and its present target is the study of
the thermal S-Z effect on a sample of nearby clusters of galaxies
in order to provide good estimates of $H_0$ and of $\Omega_M$.

\begin{figure}[htpb]
\begin{center}
\includegraphics[width= 10cm,height=14cm,angle=90]{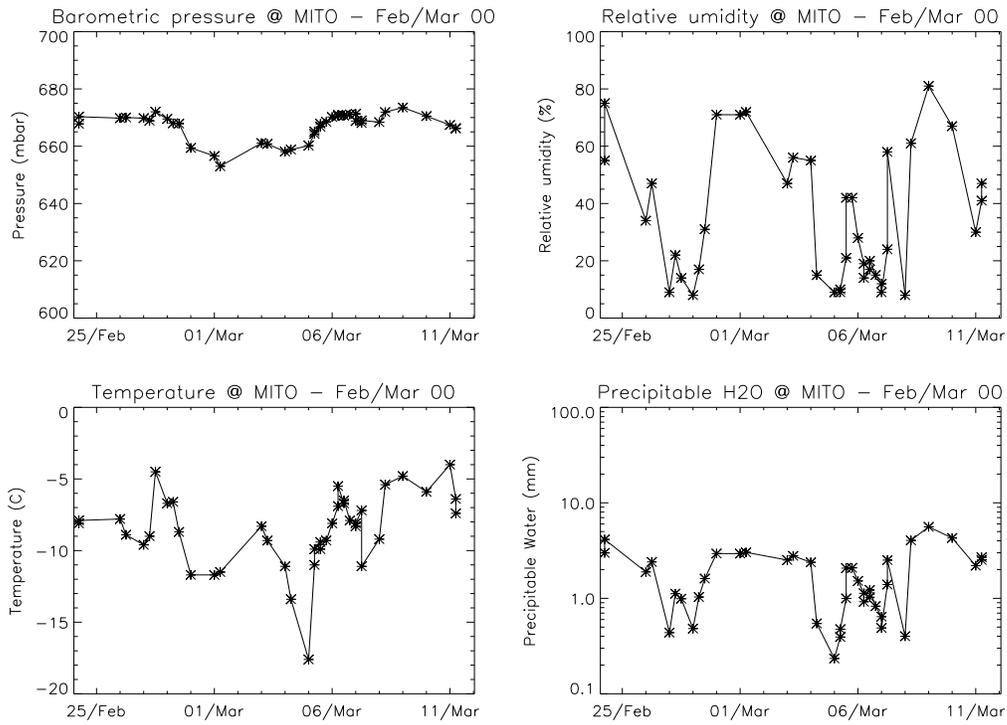}
\end{center}
\caption{Temperature, pressure, relative umidity, and calculated
pwv @ MITO during the last Feb/Mar campaign} \label{fig:meteo}
\end{figure}

In the frequency range where the S-Z distortion makes its
transition from a decrement to an increment, the atmosphere is not
very transparent and its contribution must be, therefore, limited.
The most important features of a good mm-site are then a high
atmospheric transmission and low atmospheric fluctuations. During
the last winter observational campaigns at MITO, low values of the
precipitable water vapor (pwv) content were computed, showing
atmospheric conditions as good as other important sites around the
world (see Fig.\ref{fig:meteo} and Table \ref{tab:sites}).

\subsection*{The instrument}
\label{sub:instruments}

The CMB temperature decrement is only at a level of few hundreds
microkelvin even when measured toward clusters with very large
and hot IC gas, thus observations of the S-Z effect require a
highly sensitive instrument with small and well understood
systematic errors. MITO observations are performed with a 2.6 m
aplanatic Cassegrain telescope (see Table \ref{tab:telescope})
studied for mm-differential observations. An electromechanical
system allows sky modulation in order to reduce fluctuations in
the atmospheric emission, while baffling techniques are used to
reject spurious signals due to Narcissus effect. At the telescope
focal plane is located FotoMITO, a 4-channels single pixel
photometer (17 arcmin FWHM). The channels are centered at 2.1
mm,1.4 mm, 1.1 mm and 0.85 mm, matching the atmospheric
transmission windows, to accurately measure the spectral
signature of the thermal S-Z effect. The detectors are standard
composite bolometers, cooled down to 290 mK by a two-stage cycle
$^3He-^4He$ fridge. The calibrations have been made not only with
laboratory sources but also with standard sky calibrators such as
planets and HII regions. The measured values for the optical
responsivity are reported in Table \ref{tab:channels} together
with channel characteristics.

\subsection*{Observations of the COMA cluster: preliminary results}
\label{sub:observations}

Criteria generally followed to select clusters of galaxies for the
S-Z signal detection at MITO are: good visibility
(\emph{$dec>{-6}^o$,$\forall{RA}$}) , large $ y_0\,(10^{-4}) $ and
large angular size (>5 arcmin). A list of nearby clusters
proposed for MITO observations is shown in Table
\ref{tab:clusters}; the physical parameters are deduced from X-ray
data.

Coma cluster (Abell 1656, z = 0.0235) has been the first target
in the MITO cosmological project, selected for its good
visibility, the high observational efficiency and the
high-quality of the existing X-ray measurements. Coma is the
richest nearby cluster, with relatively dense and hot IC gas, and
only a small degree of apparent ellipticity. In an initial study,
we have investigated the efficiency of detecting the S-Z effect
as a function of cluster size, field of view, and beamthrow.
Moreover, the correlation between the signal morphology and the
cluster gas physical parameters has been computed using the
\emph{$\beta$ isothermal model}, and then assuming a circular
symmetry on the sky. During the last winter campaign we have
collected about 80 drift scans, each 10 minutes long, with a
3-field sky modulation in order to integrate the source and then
to improve the signal-to-noise ratio. The residual atmospheric
fluctuations have been reduced applying a linear fit removal, a
gaussian filter and further decorrelation. A preliminary data
analysis leads us a value of the Comptonization parameter along
the line of sight through the center of the cluster equal to
$$ y_0 = (2.23\,\pm\,0.27)\,\cdot10^{-4} $$
Obviously, this estimation is a function of the cosmological
parameters involved in theoretical calculations (in particular
this value corresponds to $h_{100}=0.5$ ), and this implies a
wide range of variability of the effective thermal S-Z signal.
However, we find that with a higher value of the Hubble constant,
our result better agrees with the expected value of $y_0$ based
on the X-ray data. The full analysis procedure and final results
will be described in detail in future publications.

The first detection of the S-Z effect dates back to 1972.
Parijnskij claimed a detection toward Coma with a temperature
decrement of $1.0\,\pm\,0.5 mK$ in the radio region. After this
first attempt several efforts have been made to improve the
accuracy of measurements. A list of all the Coma measurements,
with relative results, is reported in Table \ref{tab:coma}.

\section{Conclusions}
\label{sec:conclusion}

The actual status of analysis suggests that to improve the
accuracy in the Hubble constant determination, using the S-Z
effect as a distance scale indicator, we have, first of all, to
improve the quality of the X-ray observations and then to reduce
systematic errors. For this purpose it will be essential in the
near future to develop sensitive new S-Z experiments with these
spectral and spatial characteristics: telescope and optical
elements whose effective beamsize is few arcminute (so as to
resolve the S-Z effect in nearby clusters), equipped with a
bolometer array capable of simultaneous observations at several
frequencies, both in the R-J and Wien sides. These capabilities
are now going to be achieved in the development of both
ground-based systems and stratospheric balloons.

%%%%%%%%%%%%%%%%%%%%%%REFERENCES%%%%%%%%%%%%%%%%%%%%%%%%%%%%%

%%%%%%%%%%%%%%%%%%%%%TABLES%%%%%%%%%%%%%%%%%%%%%%%
\newpage
\begin{center}
{\bf Table 1}\\
\end{center}
\vspace{0.5cm}
\begin{table}[htpb]
\centering
\begin{tabular}{|l|c|}

\hline $Site$ & $PWV\,(mm)\,-25\%$ \\
\hline Dome\,C-Antarctica\,(summer) & 0.38 \\
\hline South\,Pole-Antarctica\,(winter) & 0.19 \\
\hline Mauna\,Kea-Hawai\,(winter) & 1.05 \\
\hline Atacama-Chile\,(winter) & 0.68 \\ \hline
\end{tabular}
\vspace{0.5cm} \caption{Comparison of PWV values, relative to the
driest quartile(25\%) of the year, reported for different sites
(Lane, 1998).} \label{tab:sites}
\end{table}
%%%%%%%%%%%%%%%%%%%%%%%
\newpage
\begin{center}
{\bf Table 2}\\
\end{center}
\vspace{0.5cm}
\begin{table}[htpb]
\centering
\begin{tabular}{|l|l|}

\hline primary mirror diameter & 2600 mm \\ \hline primary vertex
curvature radius & 2494.54 mm \\ \hline primary conic constant &
-1.009 \\ \hline primary f/\# & 0.48 \\ \hline secondary mirror
diameter & 410 mm \\ \hline subreflector vertex curvature radius &
539.78 mm \\ \hline subreflector conic constant & -1.908 \\ \hline
vertex interdistance & 1018.7 mm \\ \hline minimum detector f/\# &
4.07 \\ \hline effective focal length &  8151 mm \\ \hline focal
scale ratio & 25"/mm \\ \hline

\end{tabular}
\vspace{0.5cm} \caption{MITO telescope parameters}
\label{tab:telescope}
\end{table}
%%%%%%%%%%%%%%%%%%%%%%%
\newpage
\begin{center}
{\bf Table 3}\\
\end{center}
\vspace{0.5cm}
\begin{table}[htpb]
\centering
\begin{tabular}{|l|l|l|l|l|}
\hline  & Ch1 & Ch2 & Ch3 & Ch4 \\ \hline wavelength$^\dagger$
($\mu$m)& 2000 & 1430  &  1110  &  895 \\ \hline
frequency$^\dagger$ (GHz)& 150& 209& 270& 335 \\ \hline
wavenumber$^\dagger$ (cm$^{-1}$)& 5.0 & 6.9 & 9.0 & 11.2 \\ \hline
h$\nu$/KT$_{CBR}^\dagger$ & 2.64 &3.69& 4.75& 5.89\\ \hline
bandwidth (FWHM \%) & 21 & 14 & 12 & 10
\\ \hline noise (nV/Hz$^{1/2}$) @3Hz &  10 & 11 & 8 &  27 \\ \hline
optical responsivity ($\mu{K}/nV$) & $ 430\pm40 $ & $370\pm35$ &
$400\pm40$ &  $345\pm40$ \\ \hline
\end{tabular}

\begin{center}
($^\dagger$ effective values for flat source)
\end{center}
\vspace{0.5cm} \caption{Performances of FotoMITO channels}
\label{tab:channels}
\end{table}
%%%%%%%%%%%%%%%%%%%%%%%%%%
\newpage
\begin{center}
{\bf Table 4}\\
\end{center}
\vspace{0.5cm}
\begin{table}[htpb]
\centering
\begin{tabular}{|l|l|l|l|l|l|l|}

\hline $Source$ & $z$ & $\theta_c$ & $T$ & $n_c$ & $\beta$ & $y_0$ \\
& & $arcmin$     & $keV$ & $10^{-3}cm^{-3}$ & & $10^{-4}$
\\ \hline
A401   & 0.0737 &  2.37  & $8.00^{+0.40}_{-0.40}$ &
$4.0^{+1.0}_{-1.0}$ & $0.63 \pm0.01$ &    1.08
\\ \hline A426   & 0.0179 & $9.3$& $6.79^{+0.12}_{-0.12}$ &
$3.7-5.4$  & $0.63 \pm 0.01$ & 1.16
\\ \hline
A478   & 0.0881 & $2.3$ &  $6.90^{+0.35}_{-0.35}$ &
$9.55^{+1.75}_{-1.75}$  & $0.75\pm0.01$  & 2.14\\
\hline A1367  & 0.0214 & $10.84$& $3.50^{+0.18}_{-0.18}$ &1.0 &
$0.50 \pm0.10$& 0.13\\ \hline A1656 \footnotemark & 0.0231 &
$10.5$ & $8.2^{+0.2}_{-0.2}$ & $2.89^{+0.04}_{-0.04}$ & $0.75
\pm0.03$ & 1.51 \\ \hline  A1795  & 0.0631 & $3.94$ &
$5.88^{+0.14}_{-0.14}$ & $6.2-4.1$  &$0.83 \pm0.02$ & 1.12
\\ \hline A2029  & 0.0765 & $2.39$ & $8.47^{+0.41}_{-0.36}$ &
3.8 &$0.68 \pm0.01$ & 0.9\\ \hline A2142  & 0.0899 & $3.11$&
$9.70^{+1.30}_{-1.30}$ & $6.97^{+0.41}_{-0.41}$ & $0.74 \pm0.01$
& 3.0 \\ \hline  A2163  & 0.2030 & $1.64$ &
$14.69^{+0.85}_{-0.85}$ &9.0 & $0.73 \pm0.02$ &5.57 \\
\hline A2199  & 0.0999 & $0.95$ &  $4.10^{+0.08}_{-0.08}$ &
$1.8-3.6$ & $0.64 \pm0.01$ & 0.17\\ \hline A2255 & 0.0808 &
$4.47$ & $7.30^{+1.10}_{-1.70}$ & 3.0& $0.75 \pm0.02$ &1.21\\
\hline A2256  & 0.0581 & $5.00$ & $7.51^{+0.19}_{-0.19}$
& $3.55^{+0.18}_{-0.18}$ & $0.78 \pm0.01$ & 1.23\\
\hline A2319  & 0.0559 & $4.62$ &
 $9.12^{+0.15}_{-0.15}$ &4.0 & $0.68
\pm0.05$ &1.50 \\ \hline
\end{tabular}
\vspace{0.5cm} \caption{Main features, deduced from X-ray
observations, of SZ clusters proposed for MITO}
\label{tab:clusters}
\end{table}
\footnotetext{Briel et al., 1992}
%%%%%%%%%%%%%%%%%%%%%%%%%%
\newpage
\begin{center}
{\bf Table 5}\\
\end{center}
\vspace{0.5cm}
\begin{table}[htpb]
\centering
\begin{tabular}{|c|c|c|c|c|c|l|} \hline

  $\lambda(mm)$ & $\nu(Ghz)$ & $\Delta{\nu}(GHz)$ & $FWHM(')$ &
$Beamthrow(')$ &
  $\Delta{T}_{ant}(mK)$ & Reference\\ \hline
  40.0 & 7.5 & 0.7 & $1.3\times40$ & $290 scan$ & $-1.0\pm0.5$  &
{P.}\footnotemark[1] \\ \hline
  20.0 & 15.0 & 0.4 & 2.2 & $17.4\,-\,60/120 scan$ & $+0.8\pm1.8$ &
R.\footnotemark[2] \\ \hline
  9.0 & 31.4 & 1.0 & 3.6 & 9 & $-0.19\pm0.22$ & L.\& P.\footnotemark[3]  \\
\hline
  28.3 & 10.6 & 1.1 & 4.5 & 15 & $+0.88\pm0.50$ & B. et al.\footnotemark[4]
\\ \hline
  9.4 & 32 & 5.7 & 7.35 & 22.16 & $-0.27\pm0.03$ & H. et
al.\footnotemark[5]  \\ \hline
  2.1 & 143 & 30.0 & 28 & $\pm40$ & $-0.31\pm0.39$\footnotemark[6] & S. et
al.\footnotemark[7] \\
  1.4 & 214 & 30.0 &  &  &  & \\
  1.1 & 273 & 32.7 &  &  &  & \\
  0.85 & 353 & 35.3 &  &  &  & \\ \hline

\end{tabular}
\vspace{0.5cm}
  \caption{Main features of the experiments realized to detect
  the S-Z signal toward the Coma cluster.} \label{tab:coma}
\end{table}
\footnotetext[1]{Parijnskij, 1973} \footnotetext[2]{Rudnick,
1978} \footnotetext[3]{Lake and Partridge, 1980}
\footnotetext[4]{Birkinshaw et al., 1980} \footnotetext[5]{Herbig
et al., 1995} \footnotetext[6]{see Birkinshaw, 1999}
\footnotetext[7]{Silverberg et al., 1997}
%%%%%%%%%%%%%%%%%%%%%%%%%%%%FIGURES%%%%%%%%%%%%%%%%%%%%%%%
\newpage
\begin{center}
{\bf Figure 1}\\
\end{center}
\vspace{0.5cm}
%%%%%%%%%%%%%%%%%%%%%%%%%%%%%
\newpage
\begin{center}
{\bf Figure 2}\\
\end{center}
\vspace{0.5cm}
%%%%%%%%%%%%%%%%%%%%%%%%%%%%%%

\end{document}